\documentclass[12pt,preprint]{aastex} 

\newcommand{\NHI}{\ensuremath{N({\rm H \; \mbox{\small\rm I}})}}
\newcommand{\nh}{\ensuremath{n_{\rm H}}}
\newcommand{\abund}[1]{\ensuremath{A({\rm #1})}}
\newcommand{\htwo}{\ensuremath{{\rm H}_2}}

\newcommand{\lya}{Lyman-$\alpha$}
\newcommand{\dla}{DLA}
\newcommand{\z}{$z$}

\newcommand{\percc}{cm$^{-3}$\relax}

\newcommand{\HI}{H$\;${\small\rm I}\relax}

\newcommand{\CII}{C$\;${\small\rm II}\relax}
\newcommand{\SiII}{Si$\;${\small\rm II}\relax}
\newcommand{\FeII}{Fe$\;${\small\rm II}\relax}

\newcommand{\cstar}{C$\;${\small\rm II}$^*$\relax}
\newcommand{\sistar}{Si$\;${\small\rm II}$^*$\relax}

\newcommand{\columnratio}{\ensuremath{N(\mbox{\sistar})/N(\mbox{\cstar})}}

\newcommand{\lowerfs}{\ensuremath{^2P_{1/2}}}
\newcommand{\upperfs}{\ensuremath{^2P_{3/2}}}

\newcommand{\Akj}{\ensuremath{A_{21}}}
\newcommand{\Bkj}{\ensuremath{B_{21}}}
\newcommand{\Bjk}{\ensuremath{B_{12}}}
\newcommand{\gammajk}{\ensuremath{\gamma_{12}}}
\newcommand{\gammakj}{\ensuremath{\gamma_{21}}}

\newcommand{\pss}{PSS~1443+27}
\newcommand{\maxtemp}{954}
\newcommand{\alphamaxtemp}{524}
\newcommand{\solarmaxtemp}{954}

\begin{document}

\title{Cold Neutral Gas in a $z=4.2$ Damped
  Lyman-$\alpha$ System: Fuel for Star Formation}

\author{J. Christopher Howk\altaffilmark{1}, Arthur M.
  Wolfe\altaffilmark{1}, Jason X. Prochaska\altaffilmark{2,3}}

\altaffiltext{1}{Center for Astrophysics \& Space Sciences, University
  of California, San Diego, La Jolla, CA 92093 }

\altaffiltext{2}{University of California Observatories, Lick
  Observatory, UC Santa Cruz, Santa Cruz CA 95064}

\altaffiltext{3}{Visiting Astronomer, W. M. Keck Telescope. The Keck
  Observatory is a joint facility of the University of California and
  the California Institute of Technology.}


\begin{abstract}
  
  We discuss interstellar temperature determinations using the
  excitation equilibrium of the $^2P$ levels of \SiII\ and \CII.  We
  show how observations of the \upperfs\ fine structure levels of
  \SiII\ and \CII\ (which have significantly different excitation
  energies, corresponding to $\sim413$ and 92 K, respectively) can be
  used to limit gas kinetic temperatures.  We apply this method to the
  $z=4.224$ damped Lyman-$\alpha$ system toward the quasar \pss.  The
  lack of significant absorption out of the \SiII\ \upperfs\ level and
  the presence of very strong \CII\ \upperfs\ provides an upper limit
  to the temperature of the \cstar -bearing gas in this system.
  Assuming a solar Si/C ratio, the observations imply a $2\sigma$
  limit $T<\maxtemp$ K for this absorber; a super-solar Si/C ratio
  gives stricter limits, $T<\alphamaxtemp$ K.  The observations
  suggest the presence of a cold neutral medium; such cold gas may
  serve as the fuel for star formation in this young galaxy.

\end{abstract}

\keywords{ISM: evolution -- ISM: atoms -- galaxies: evolution --
  quasars: absorption lines}


\section{Introduction}

High-redshift damped \lya\ systems (\dla s) are the highest column
density class of QSO absorption lines.  Defined by $\log N(\mbox{\HI})
\ge 20.3$ (Wolfe et al. 1986), these systems are thought to trace the
interstellar medium (ISM) of high-redshift galaxies.  Dedicated
surveys over the past two decades have helped trace the global
properties of high-redshift DLAs, including their contribution to the
cosmological baryon density (Storrie-Lombardi \& Wolfe 2000; Prochaska
\& Herbert-Fort 2004), their chemical enrichment (e.g., Prochaska et
al. 2003), their dust content (e.g., Pettini et al.\ 1994), and
molecular fraction (Ledoux, Petitjean, \& Srianand 2003).  These
studies have demonstrated that the DLAs have a baryonic mass density
comparable to the mass density of modern galaxy disks, that the
metallicity of high-$z$ DLAs is slowly increasing, and that the
majority of DLA sight lines have low dust-to-gas ratios and molecular
fractions.

Understanding the detailed physics of the ISM in DLAs is an important
step in understanding high-redshift galaxies in general.  Wolfe,
Prochaska, \& Gawiser (2003a) have constructed detailed models for the
thermal equilibrium of the ISM in a set of DLAs, calculating the
heating rate experienced by the gas due to the ultraviolet emission
from young hot stars.  They used observations of absorption out of the
\upperfs\ level of \CII\ (hereafter \cstar) -- a direct indicator of
the cooling rate through [\CII] 158 $\mu$m emission (Pottasch,
Wesselius, van Duinen 1979) -- to infer the actual heating rate
experienced by the gas (assuming thermal equilibrium).  Their
comparison of the inferred and calculated heating rates suggests DLAs
harbor significant star formation.

The models of Wolfe et al. (2003a,b) required that the observed
\cstar\ in DLAs arise in a cold neutral medium (CNM), i.e., in gas
with temperatures $T\la1000$ K; their WNM models give SFRs that
violate observations of the bolometric background.  The detection of
\htwo\ absorption in some high-\z\ DLAs is further evidence that at
least some of these systems contain a CNM (Ledoux et al. 2003;
Hirashita \& Ferrara 2005).  However, such temperatures are at odds
with 21-cm absorption studies; all $z\ga3$ DLAs searched for 21-cm
absorption show $T_S\ga1400$\,K ($2\sigma$) (Kanekar \& Chengalur
2003).  Wolfe et al.  (2003b) argue that this discrepancy is likely
due to the differing properties of the sight lines probed by the
optical background sources and by the more extended radio sources.


In this Letter we present a method for determining the kinetic
temperature of interstellar matter based solely on basic atomic
physics.  Our method compares the excitation of the upper \upperfs\ 
fine-structure levels in \SiII\ and \CII, which have excitation
energies that differ by a factor of four.  We describe our approach in
\S \ref{sec:method}.  We apply this technique to the
$z_{abs}\approx4.224$ DLA toward the quasar \pss\ in \S
\ref{sec:psscnm}, demonstrating that this DLA contains a substantial
reservoir of cold gas.  Lastly, we discuss the implications of this
temperature determination in \S \ref{sec:discussion}.

\section{Determining  Gas Temperatures from $N(\mbox{\sistar})$ 
  and $N(\mbox{\cstar})$}
\label{sec:method}

In this section we discuss how analysis of the $^2P$ fine structure
excitation in \SiII\ and \CII\ can be used to limit interstellar
kinetic temperatures.  Srianand \& Petitjean (2000) have used a
similar analysis to limit temperatures and densities in an associated
absorber.  Silva \& Viegas (2002, hereafter SV) give a detailed
summary of the excitation equilibrium of fine structure lines in DLAs.

We treat the excitation of the $^2P$ fine structure states as a
two-level atom.\footnote{This is appropriate given the temperatures of
  the ISM probed by low-ionization absorption lines ($T\ll30,000$ K).}
The equilibrium ratio of the densities in the upper and lower levels
(Spitzer 1978), $n_2$ and $n_1$, respectively, is
\begin{equation}
  \frac{n_2}{n_1} = 
  \frac{\Bjk u_{\nu_{12}}(z) + \Gamma_{12} + \sum\limits_k n_k \gammajk^k}
  {\Akj + \Bkj u_{\nu_{12}}(z) + \Gamma_{21} + \sum\limits_k n_k \gammakj^k}.
\label{eqn:populations}
\end{equation}
The quantities \Akj, \Bkj, and \Bjk\ are the familiar Einstein
transition probabilities.  The energy density of the cosmic microwave
background, $u_{\nu_{12}}(z)$, for direct excitation of the
transitions is calculated assuming a standard cosmology, i.e.,
$T_{CMB}=T_0(1+z)$, where $T_0=2.725$ K (Mather et al.  1999).  The
fluorescent rates $\Gamma_{12} = \Gamma_{21} = 0$ due to the great
opacity of the ground-state transitions of interest (Sarazin, Rybicki,
\& Flannery 1979; see Wolfe et al.  2003a).  The summations describe
excitations and deexcitations with collision partners $k$($=e^-$,
$p^+$, and H$^0$), where $n_k$ is the particle density of each partner
and $\gammajk^k$ and $\gammakj^k$ are the Maxwellian-averaged
collision rate coefficients for excitation and deexcitation.  The
collision rates are related to one another and to $\Omega_{12}^k(T)$,
the collision strength (Spitzer 1978), by:
\begin{equation}
  \gammajk^k \propto \gammakj^k \exp{(-kT_{12}/kT)} \\
  \propto \Omega_{12}^k(T)\, T^{-1/2} \exp{(-kT_{12}/kT)},
\label{eqn:gammadefinition}
\end{equation}
where $kT_{12}$ is the energy of the \upperfs\ level above the ground
state, and $T$ is the gas kinetic temperature.  We adopt atomic data
from the same sources as SV throughout.

The column density of material in the \upperfs\ level of \CII\ is
\begin{equation}
N(\mbox{\cstar}) \approx 
    \int \frac{n_{\rm C\; II^*}}{n_{\rm C\; II}} \abund{C} \nh \, ds,
\label{eqn:cstarcolumn}
\end{equation}
where $\abund{C}$ is the gas-phase abundance of carbon, \nh\ is the
density of neutral hydrogen, $ds$ is the differential pathlength along
the line of sight, and the ratio $n_{\rm C\; II^*}/n_{\rm C\; II}$ is
a function of four quantities: $z_{abs}$, \nh, $x\equiv n_e/\nh
\approx n_p/\nh$, and $T$ (Eqn.  \ref{eqn:populations}).  We have
assumed $n_{\rm C\; II} \approx n_{\rm C} \equiv \abund{C} \nh$,
consistent with limits on \ion{C}{1} in DLAs (see Figure 12 of Wolfe
et al.  2003b).  A similar expression applies for \sistar.  The column
densities of \cstar\ and \sistar\ are therefore density- and
excitation-weighted integrals over pathlength.

\begin{figure}
\epsscale{0.9}
\plotone{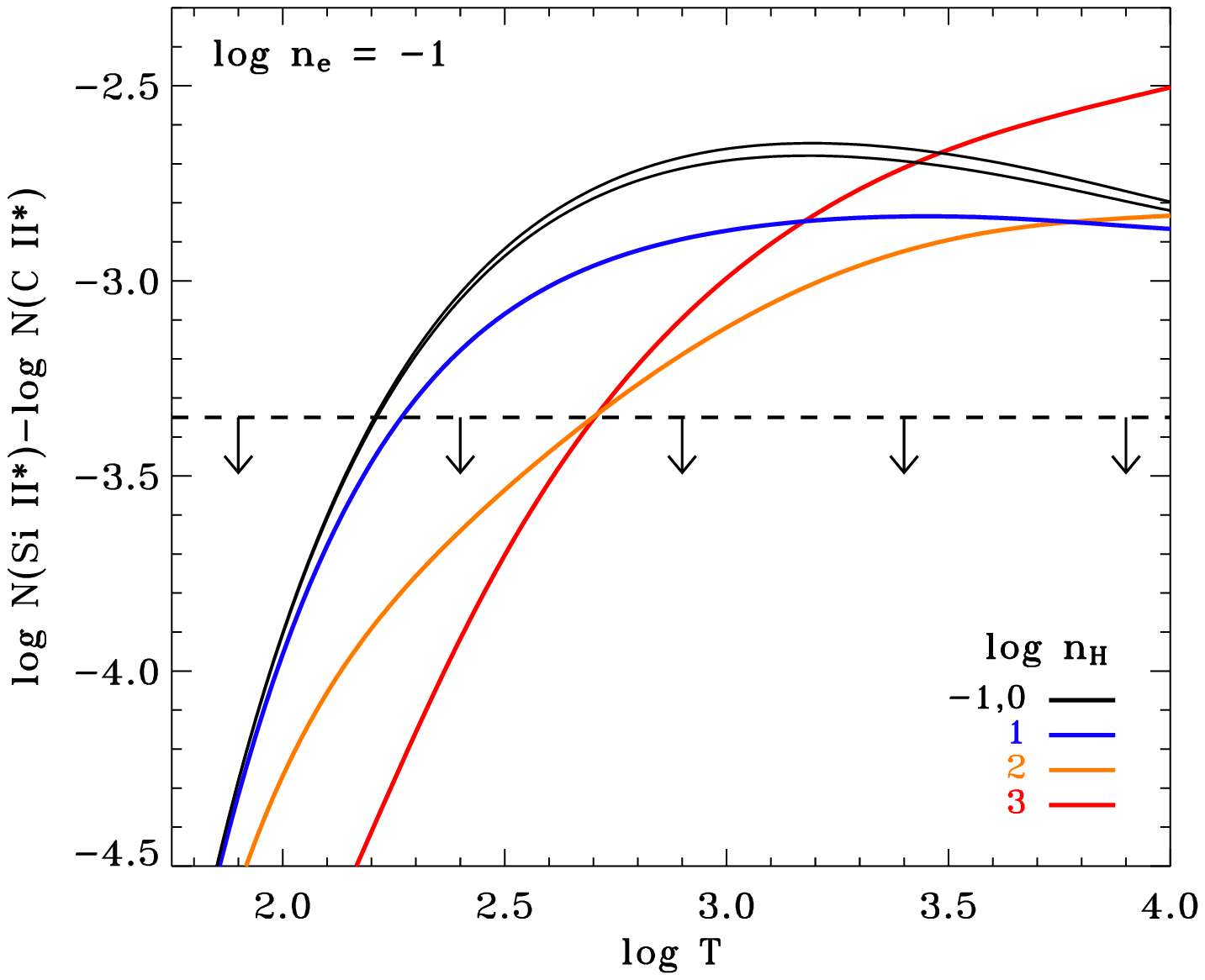}
\caption{Predicted \columnratio\ for a DLA at $z\sim4.22$ with 
  $\log n_e=-1$ and a large range of hydrogen densities.  The dashed
  line represents the $2\sigma$ upper limits to \columnratio\ for the
  DLA toward \pss.  Electrons generally dominate the collisional
  excitation of \SiII\ and \CII\ ($\gammajk^e \gg \gammajk^{\rm H}$).
  Although at very low values of $x \equiv n_e/\nh$ (bottom curves),
  \HI\ collisions become important.  The change in slope of the
  lowest-$x$ curves is due to the larger relative importance of \HI\ 
  in collisional excitations of \CII\ compared with \SiII\ (compare
  Figures 3 and 7 of SV) and the differing temperature dependence for
  $\gammajk^{\rm H}$.  The ratios of the \upperfs\ to \lowerfs\ levels
  of \CII\ and \SiII\ can be used to constrain \nh\ and $x$.}
\label{fig:ratio}
\end{figure}

To understand the usefulness of \sistar\ and \cstar\ to limit gas
temperatures, imagine that the excitation of the \upperfs\ states were
only due to electron collisions and deexcitation due to spontaneous
emission.  In this case $N(\mbox{\sistar})/N(\mbox{\cstar}) \propto
[\abund{Si}/\abund{C}] \times [\Omega_{12}^e(T)_{\rm
  Si}/\Omega_{12}^e(T)_{\rm C}] \times \exp\{-(T_{12}^{\rm Si} -
T_{12}^{\rm C})/T\}$, i.e., the ratio is only a function of
temperature and the Si/C ratio.
%
%
Figure \ref{fig:ratio} shows the ratio \columnratio\ as a function of
temperature, including all of the terms in Eqn.
(\ref{eqn:populations}), for several densities at $z\sim4.2$ assuming
[Si/C]=0.\footnote{[X/Y]$\, \equiv \log N(X)/N(Y) - \log
  (X/Y)_\odot$.}  $N(\mbox{\sistar})/N(\mbox{\cstar})$ varies slowly
with temperature for $T\ga1,000$ K (for all but the highest densities)
and more strongly at lower temperatures.  The sensitivity to density
is modest, especially for low $x$, because the populations of the
\upperfs\ levels of \SiII\ and \CII\ depend on density in a similar
manner.  We use the ratio of the \upperfs\ to \lowerfs\ levels of
\CII\ and \SiII\ to constrain the allowable range of densities.

Gas at $T\la1,000$ K produces ratios $\log N(\mbox{\sistar}) /
N(\mbox{\cstar}) \la -2.8$; the corresponding optical depth ratio of
the 1264.738 \AA\ line of \sistar\ to the typically-unresolved
1335.663 and 1335.708 \AA\ doublet of \cstar\ is
$\tau(1264)/\tau(1335) \la 0.015$ ($f$-values from Morton 2003).
Thus, the use of this ratio to distinguish cool from warm temperatures
requires large \cstar\ optical depths and good S/N at \sistar.
Observations of \columnratio\ will allow a range of parameters $\nh,
x$, and $T$, which apply to the \cstar -bearing gas.  However, the
maximum allowable temperature can be low if the limits on
\columnratio\ are low.

\section{Application to the $z=4.224$ Damped
  Lyman-$\alpha$ System Toward \pss}
\label{sec:psscnm}

Prochaska et al. (2001) present high-resolution ($R\sim40,000$), high
S/N observations of the $z=4.224$ \dla\ toward of the quasar \pss\ 
using HIRES (Vogt et al. 1994) on the Keck I 10-m telescope.  The
properties of this DLA are summarized in Table \ref{tab:properties}.
Due to its redshift and relatively strong metal lines, column
densities can only be measured for a few species.  Table
\ref{tab:properties} gives a measurement of \FeII\ (from the 1611.2
\AA\ transition) and meaningful limits on \cstar, \SiII\, and \sistar.
All but the last of these are from Prochaska et al. (2001).

\begin{figure}
\epsscale{0.9}
\plotone{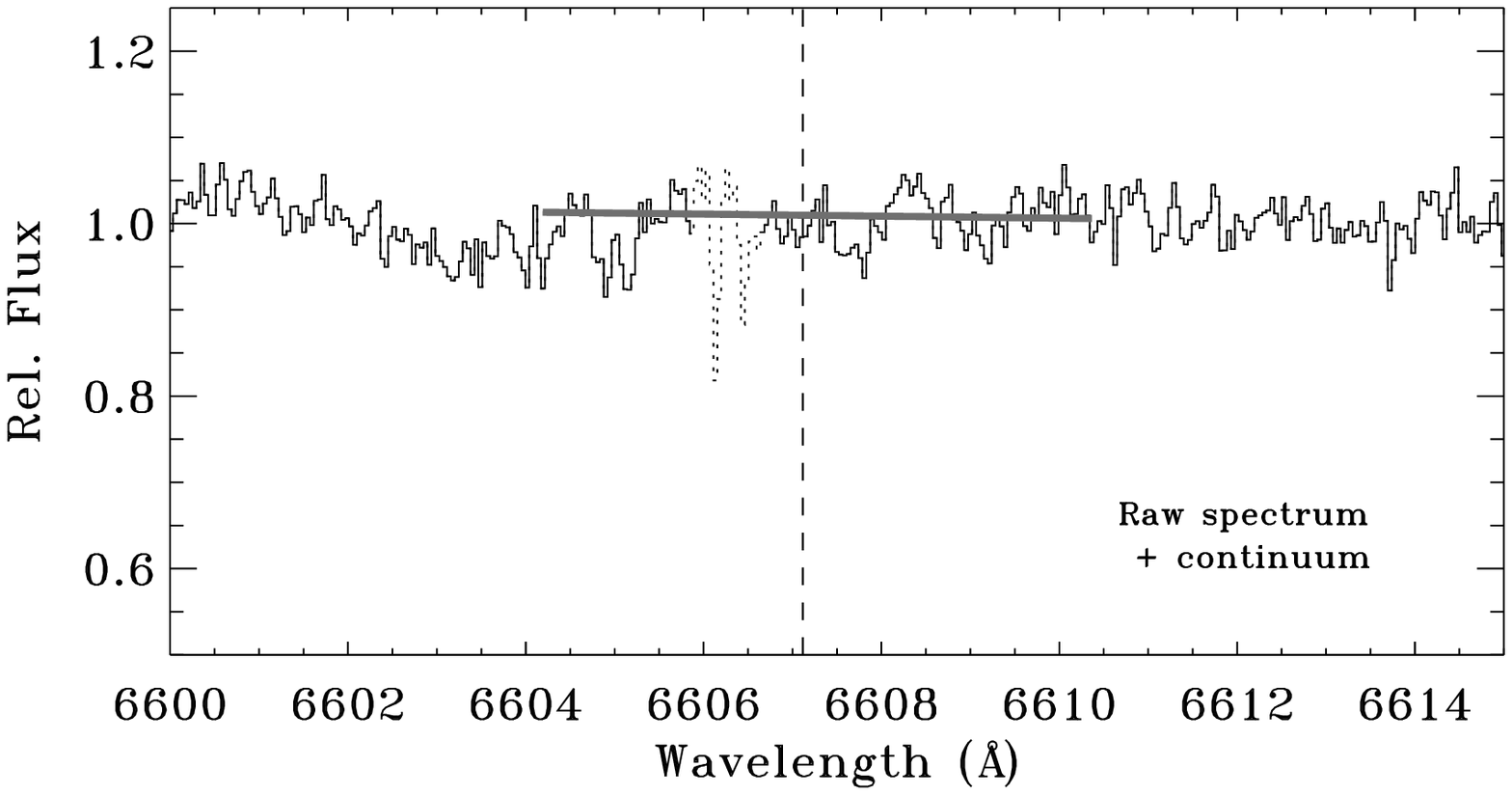}
\caption{Observed spectrum near the expected position of  
  \sistar\ for the $z=4.224$ DLA toward \pss\ (shown by the dashed
  line).  The thick grey line represents our fitted continuum.  The
  dotted region denotes the location of two poorly-subtracted sky
  emission lines, which cause the narrow depressions in this region of
  the spectrum.
  \label{fig:continuum}}
\end{figure}

\begin{figure}
\epsscale{0.75}
\plotone{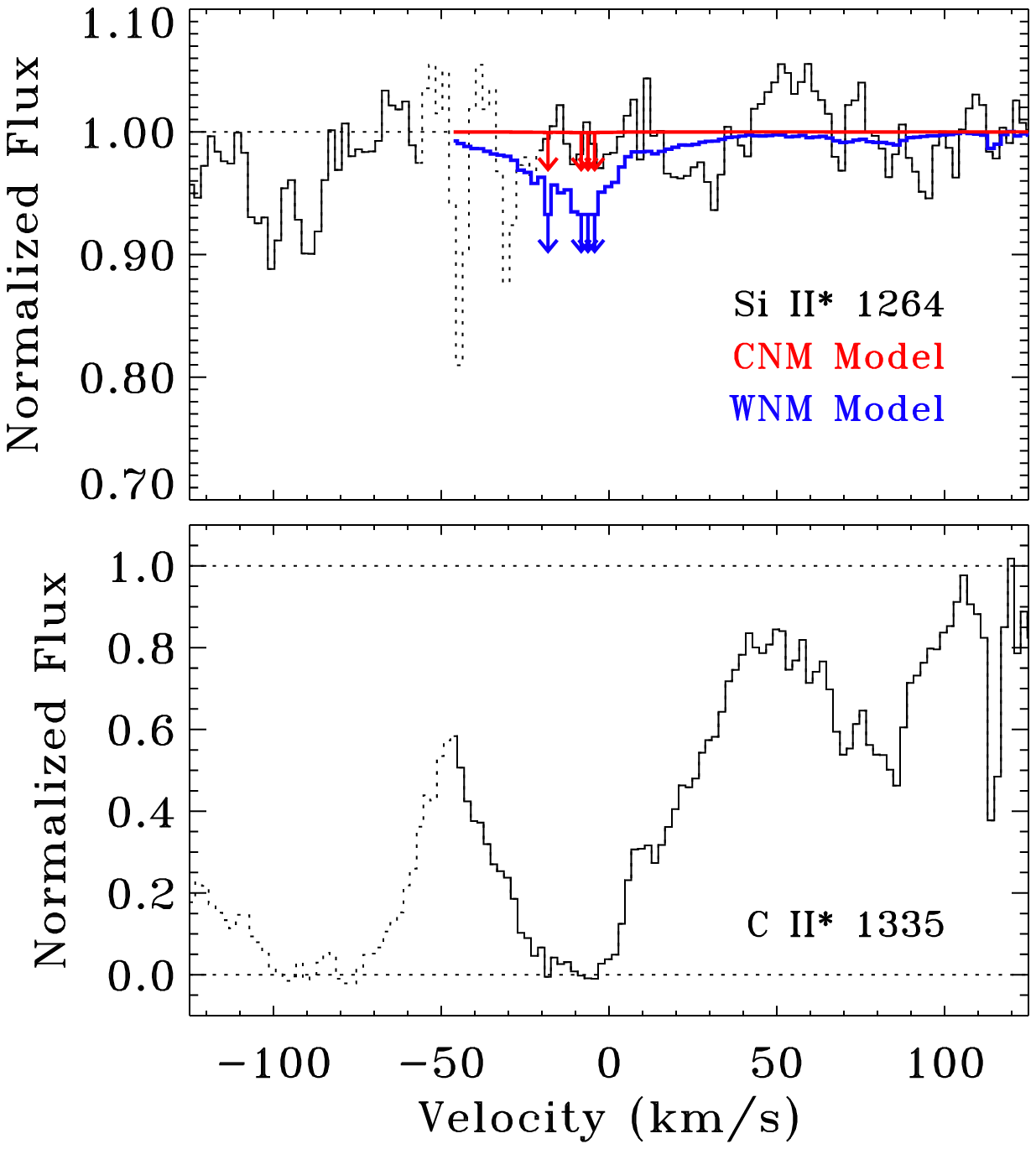}
\caption{Normalized \sistar\ and \cstar\ profiles for the $z=4.224$ 
  DLA toward \pss.  Included in the panel showing \sistar\ are two
  predicted \sistar\ profiles derived from the observed \cstar\ 
  profile assuming standard WNM ({\em blue}, with $\nh=1$ \percc,
  $x\equiv n_e/\nh = 0.1$, and $T=8,000$ K) and CNM ({\em red}; with
  $\nh=10$ \percc, $x=10^{-3}$, and $T=80$ K) properties.  These
  models assume [Si/C]$\, =0$.  The arrows denote regions of definite
  saturation in the \cstar\ profile.  The dotted region in the
  \sistar\ profile denotes the location of two poorly-subtracted sky
  emission lines.
  \label{fig:profile}}
\end{figure}

Figure \ref{fig:continuum} shows the HIRES spectrum at the expected
location of \sistar\ in this absorber, including our fitted continuum.
Our limit to \sistar\ was determined using an empirically-measured
signal-to-noise ratio (following Sembach \& Savage 1992) and includes
continuum placement uncertainties.  Figure \ref{fig:profile} shows the
normalized profiles of \sistar\ and \cstar.  Also shown are models for
the expected \sistar\ absorption for a canonical warm neutral medium
(WNM; with $\nh=1$ \percc, $x = 0.1$, and $T=8,000$ K) and CNM (with
$\nh=10$ \percc, $x=10^{-3}$, and $T=80$ K).  The models are derived
by scaling the \cstar\ profile by the amount predicted using Eqns.
(\ref{eqn:populations}) and (\ref{eqn:cstarcolumn}) and assuming
[Si/C]$\,=0$.  The hypothesis that \cstar\ absorption toward \pss\ 
arises in a canonical WNM is not consistent with the observations.

To limit $T$, we calculate the excitation balance of the fine
structure levels of \SiII\ and \CII\ for densities $-4 \le \log n_e
\le +2$ and $-2 \le \log \mbox{\nh} \le +3$ and temperatures $1.5 \le
\log T \le 4.5$.  We compare these calculations with our observational
limits for \columnratio, $N(\mbox{\cstar}) / N(\mbox{\CII})$, and
$N(\mbox{\sistar}) / N(\mbox{\SiII})$ to constrain the physical
properties (\nh, $x$, and $T$) of the gas in the \cstar -bearing
gas.\footnote{These three physical quantities are coupled in the
  calculations.  A wide range of T-dependent $x$ and $\nh$ values are
  allowable, and only T is stringently constrained in our analysis.}

The allowable range of physical conditions in the \pss\ DLA then
depends upon the ratios of Si/C and C/Fe, which are used for
determining the expected \columnratio\ and $N(\mbox{\cstar}) /
N(\mbox{\CII})$, respectively.  \CII\ is rarely measurable in DLAs due
to saturation effects; we estimate $N(\mbox{\CII})$ from
$N(\mbox{\FeII})$ (Table \ref{tab:properties}) with an assumed
abundance ratio.  As a fiducial, we adopt the solar system ratios from
Grevesse \& Sauval (1998): $\log ({\rm C/Fe})_\odot = +1.02$ and $\log
({\rm Si/C})_\odot = -0.96$.  This gives $\log N(\mbox{\CII}) =
16.22\pm0.06$.  Nucleosynthetic effects are unlikely to modify the
C/Fe ratio, as a roughly solar value is found in low-metallicity Milky
Way stars (e.g., Carretta, Gratton, \& Sneden 2000).  However,
nucleosynthetic effects could enhance $\alpha$ element abundances,
leading to high Si/C ratios (see, e.g., Prochaska 2004).  In one case
where the \CII\ measurements may be reliable, D'Odorico \& Molaro
(2004) find $\log N(\mbox{\SiII})/N(\mbox{\CII}) = -0.74\pm0.07$ and
$\log N(\mbox{\CII})/N(\mbox{\FeII}) = +1.25\pm0.12$ (assuming the
\FeII\ column from Lu et al.  1996).  This implies [C/Fe]$\,=+0.23$
and [Si/C]$=+0.22$ compared with our adopted solar system abundances.
The overall metallicity of this absorber is [Si/H]$\, = -1.72$,
however, which is significantly lower than the metallicity of the DLA
studied in this work.

The presence of dust will raise the C/Fe ratio due to differential
depletion.  The depletion of Fe in DLAs is typically modest, similar
to the ``halo'' and ``disk-halo'' clouds in the Milky Way (Savage \&
Sembach 1996) with [Fe/H]$\ga -0.9$.\footnote{The intrinsic gas+dust
  abundance in the Milky Way is [Fe/H]$\approx0$.  Thus, unlike the
  case in DLAs, the sub-solar gas-phase Fe/H quoted here is due
  entirely to depletion into dust grains.}  For diffuse clouds in the
Milky Way, Sofia et al.  (2004) find [C/H]$\sim-0.3$.  Thus, we expect
[C/Fe]$\la +0.6$.  In low-depletion Milky Way clouds, and by extension
DLAs, Si is likely not depleted much differently from C, and we assume
[Si/C]$= 0$.  The small number of species for which measurements are
possible in the DLA toward \pss\ make it difficult to judge the degree
of Fe depletion.  Moderate depletion is suggested by [Si/Fe]$\, >
+0.17$, but this could also result from $\alpha$-enhanced abundances.


We give temperature limits for the \cstar -bearing gas in the
$z\sim4.224$ \dla\ toward \pss\ for six different abundance
assumptions in Table \ref{tab:results}.  The first three assume
[Si/C]$\,=0$ with [C/Fe]$\, =+0.0$, +0.3, and +0.6, allowing for
modest levels of depletion.  The next three assume [Si/C]$\, = +0.3$,
i.e., $\alpha$-enhanced abundances (see Prochaska 2004).  We note that
the model in which [Si/C]$=$[C/Fe]$=0$ is not self-consistent: scaling
both \SiII\ and \FeII\ by solar abundances gives inconsistent values
for the \CII\ column density.

The highest limit is $T<\solarmaxtemp$ K, assuming assuming [C/Fe]$\,
=+0.6$ and [Si/C]$\, =0$.  This is likely an extreme assumption for
the C/Fe ratio.  The highest temperature limit for $\alpha$-enhanced
abundances, which we prefer due to the distribution of [Si/Fe] at very
low metallicities (where dust depletion effects are likely small; see
Prochaska 2004), is $T<\alphamaxtemp$ K.  Temperatures above 1400 K,
consistent with $T_s$ limits for the $z\ge2.9$ DLAs observed at 21-cm
(Kanekar \& Chengalur 2003), require [C/Fe]$\,\ga+0.75$ to +0.8 for
both solar and $\alpha$-enhanced abundances.  We feel this is
inconsistent with our knowledge of differential depletion in the Milky
Way and DLAs.


\section{Discussion}
\label{sec:discussion}

We have presented a method for limiting gas temperatures in the ISM of
galaxies through measurements and analysis of the \upperfs\ fine
structure levels of \SiII\ and \CII.  We have applied this method to
limit the properties of the \cstar -bearing gas in the $z=4.224$ DLA
toward \pss.  We rule out the hypothesis that the \cstar\ absorption
arises in a WNM.  Our conservative temperature limit for this gas is
$T \la \maxtemp$\,K; we obtain stricter temperature limits if
[Si/C]$\, > 0$ or [C/Fe]$\, \approx 0$: $T\la \alphamaxtemp$ K.
%
%
The detection of a CNM (Wolfire et al. 1995) in a high-redshift \dla\ 
is significant: while the gas seen in this DLA is likely not
associated with the dense star-forming clouds, our result demonstrates
that the physical conditions of the ISM in this system do not preclude
the existence of cold material, including, in principle, the dense
clouds from which stars could form.

The detection of a CNM in the z=4.224 absorber toward PSS 1443+27 is
the first detection at such a large redshift.  We stress that our
measurements allow a WNM as part of a multiphase medium toward \pss,
but the majority of the \cstar\ cannot come from warm
material.\footnote{We have tested the effects of multiphase absorbers
  on our technique, with as much as half of the ground-state ions
  arising in a WNM.  We find that there is very little impact on the
  derived maximum temperatures (Howk et al.  2005).}  The existence of
CNM material may be a feature of many high-\z\ DLAs, as suggested by
Wolfe et al. (2003a,b).  While $z \la 2$ \HI\ 21-cm absorption-line
measurements suggest the presence of cold \HI\ in DLAs, no 21-cm
absorption has been found in DLAs with $z\ga3$ (Kanekar \&
Chengalur).\footnote{Only six $z>2.9$ DLAs have been observed; \pss\ 
  has not been searched for 21-cm absorption.}  We note, however, that
at least three $z>2.5$ DLAs show \htwo\ absorption (Ledoux et al.
2003), indicating the presence of cold gas (Hirashita \& Ferrara
2005).  Searching for \htwo\ toward \pss\ might be difficult due to
the strength of the \lya\ forest at $z\sim4$.  In the future we will
apply our method for determining gas kinetic temperatures to DLAs for
which 21-cm and \htwo\ measurements have been attempted.

We note that the DLA toward \pss\ may not be typical.  It shows a high
metallicity ([Si/H]$\ga -1$), especially compared with other $z>4$
DLAs (Prochaska et al. 2003).  Furthermore, we chose this DLA for this
experiment because it has very strong \cstar.  It has the highest
\cstar /\HI\ ratio measured, some $+0.6$ dex above the next highest
(the $z=1.92$ DLA toward Q2206-19; Wolfe et al.  2004).  The intensity
of radiation in this DLA calculated following Wolfe et al. (2004) is
higher than all other systems.  Although the fuel for star formation
may be present in this DLA, no optical counterpart to this DLA has
been identified in deep ground-based and {\em Hubble Space Telescope}
images of \pss\ (Prochaska et al.\ 2002; L.  Storrie-Lombardi, private
communication).

The method presented here will be discussed further, and temperature
limits given for a larger number of DLAs, in Howk et al. (2005).  We
note that the use of the relative populations of the \upperfs\ levels
of \SiII\ and \CII\ may also be useful for constraining temperatures
in Milky Way gas.

\acknowledgements

The authors wish to recognize and acknowledge the very significant
cultural role and reverence that the summit of Mauna Kea has always
had within the indigenous Hawaiian community.  We are most fortunate
to have the opportunity to conduct observations from this mountain.
We thank an anonymous referee for comments that improved this work.
JCH and AMW recognize support from proposal number HST-AR-09931.01-A,
provided by NASA through a grant from the Space Telescope Science
Institute, which is operated by the Association of Universities for
Research in Astronomy, Incorporated, under NASA contract NAS5-26555.
AMW and JXP are supported by NSF grant AST 03-07824.




\begin{deluxetable}{lr}
  \tablenum{1} \tablecolumns{2} \tablewidth{0pt} \tablecaption{\pss\ 
    Damped Lyman-$\alpha$ System Properties\tablenotemark{a}
    \label{tab:properties}} \tablehead{ \colhead{Quantity} &
    \colhead{Value}} \startdata
$z_{abs}$      & 4.2240 \\
$\log N(\mbox{\HI})$ & $20.80\pm0.10$ \\
$[{\rm Fe/H}]$  & $-1.10\pm0.11$ \\
$\log N(\mbox{\ion{Si}{2}})$ & $>15.43$ \\
$\log N(\mbox{\ion{Fe}{2}})$ & $15.20\pm0.06$ \\
%
%
$\log N(\mbox{\ion{C}{1}})$  & $13.37\pm0.09$ \\
$\log N(\mbox{\ion{C}{2}}^*)$  & $>14.71$ \\
$\log N(\mbox{\ion{Si}{2}}^*)$ & $<11.24 \, (2\sigma)$ \\
$\log N(\mbox{\ion{Si}{2}}^*)/N(\mbox{\ion{C}{2}}^*)$ & $<-3.47 \, (2\sigma)$ \\
\enddata
\tablenotetext{a}{\NHI, [Fe/H] and column densities, with the
exception of \sistar, from Prochaska et al. 2001.}
\end{deluxetable}


\pagebreak

\begin{deluxetable}{ccccc}
\tablenum{2}
\tablecolumns{5}
\tablewidth{0pt}
\tablecaption{Temperature Limits for $z=4.22$ 
        Damped System\label{tab:results}}
\tablehead{
\colhead{Model} & 
\colhead{[Si/C]} &
\colhead{[C/Fe]} &
\colhead{$T$ [K]} &
\colhead{$\log n_{\rm H}$\tablenotemark{a}}}
\startdata
1 & +0.0 & +0.0 & $<478$  & 1.44 \\
2 & +0.0 & +0.3 & $<724$  & 1.04 \\
3 & +0.0 & +0.6 & $<954$  & 0.56 \\
4 & +0.3 & +0.0 & $<416$  & 1.46 \\
5 & +0.3 & +0.3 & $<436$  & 1.14 \\
6 & +0.3 & +0.6 & $<524$  & 0.64 \\
\enddata 
%
\tablenotetext{a}{Hydrogen density giving maximum allowed
  temperature in each model.}
\end{deluxetable}

\pagebreak

\end{document}